\documentclass[11pt]{article}
\usepackage{moriond,epsfig}

\def\Journal#1#2#3#4{{#1} {\bf #2}, #3 (#4)}

\def\PRL{\em Phys. Rev. Lett.}
\def\PRD{{\em Phys. Rev.} D}

\def\be{\begin{equation}}
\def\ee{\end{equation}}
\def\bea{\begin{eqnarray}}
\def\eea{\end{eqnarray}}

\def \to   {\rightarrow}
\def \pmm  {$K^0_L    \rightarrow \pi^0\mu^+\mu^-$}
\def \park {$\Sigma^+ \rightarrow p\mu^+\mu^-$}
\def \hyp  {$\Xi^0    \rightarrow \Sigma^+\mu^-\nu$}
\def \Vus  {$|V_{us}|$}
\def \JP   {$(J)^P$}
\def \Qno#1#2  {$(#1)^{#2}$}
\def \Ke  {$K^0_L \to \pi^0 e^\pm\nu$}
\def \Km  {$K^0_L \to \pi^0\mu^\pm\nu$}
\def \pccc  {$K^0_L \to \pi^+\pi^-\pi^0$}
\def \pnnn  {$K^0_L \to 3\pi^0$}
\def \cc   {$K^0_L \to \pi^+\pi^-$}
\def \nn   {$K^0_L \to 2\pi^0$}
\def \EUR  {{\em Eur. Phys. J.} C}

\begin{document}
\vspace*{4cm}

\title{RECENT RESULTS FROM KTEV}
\author{ L. BELLANTONI\\for the KTEV Collaboration }
\address{Fermi National Accelerator Laboratory,\\Batavia IL   60510, U.S.A.}
\maketitle\abstracts{The implications of the published KTeV \pmm~ result for interpreting recent \park~ results are discussed.  The status of the KTeV 
\hyp~ analysis is given.  The KTeV \Vus~ result is also given.}

\section{\pmm}
\label{sec:one}

The decay \pmm~ has three contributions to the amplitudes: a $CP$ conserving 
term from intermediate states with two photons; a $CP$ violating term from 
indirect $CP$ violation; and a direct $CP$ violating term from 2nd-order 
electroweak penguin and box diagrams.  The process continues to be of 
theoretical interest; see in particular the presentation by Christopher Smith 
at this conference.  There is also interest on the experimental side as a 
consequence of a recent unusual result from the HyperCP 
collaboration~\cite{parkPRL}.  

In searching for \park, HyperCP found three events.  This is not so unusual; 
what is peculiar is that to within the $\sim0.5\,MeV$ resolution of the mass measurements, all three events have the same dimuon mass of 214.3$\, MeV$.  
In the standard model, the muon pair is produced by an off-shell 
photon and a spread of dimuon masses is expected.  The HyperCP collaboration 
estimates that the probability of three events from an intermediate photon 
having the same mass to this level of precision is about 0.8\%, and they 
suggest that there may be a new intermediate neutral state causing this 
anomaly.

This postulated new state would be a flavor changing neutral current when 
coupling to quarks, but not when coupling to leptons.  The scale of the 
partial width for the three observed events, 
$\Gamma(\Sigma^+ \rightarrow pP^0,P^0 \rightarrow \mu^+\mu^-) 
 \sim 2.6 \times 10^{-19}\,MeV$ is too small for a strong interaction, which
have widths on the order of a few $10^{-12}\,MeV$ in $\Sigma^+$ decays.  
Also of course, the active state of research into new bound QCD states not
withstanding, a new narrow hadronic resonance in this mass range would be a 
surprise.  So we consider new point-like particles.  Allowing that the new 
interaction conserves parity and angular momentum, the only possible 
\JP~ values for decays into $\mu^+\mu^-$ are \Qno{0}{-}~ and \Qno{1}{-} .  
However if this $P^0$ is a vector current then it will also contribute to 
\pmm decay.

The existing KTeV limit~\cite{sadaPRL} on Br(\pmm) of 
$3.8 \times 10^{-10}$ at the 90\% C.L., while an order of magnitude above the 
standard model prediction~\cite{chris} of $(1.5 \pm0.3) \times 10^{-11}$, does
correspond to a partial width of
$\Gamma(K_L^0 \rightarrow \pi^0P^0,P^0 \rightarrow \mu^+\mu^-) 
 \sim 4.8 \times 10^{-24}\,MeV$, nearly 5 orders of magnitude less than
the postulated HyperCP rate.  The vector current hypothesis is thus disfavored.

Work on a new limit for Br(\pmm) based on the full KTeV data set is in
progress.

\section{\hyp}
\label{sec:two}

The first observation~\cite{ashkan} of the decay \hyp~ was made on the 1997 KTeV
dataset; here we report results based on the 1999 data, which corresponds to about
$3 \times 10^{8}$~ $\Xi^0$ decays.  We reconstruct the $\Sigma^+$ hyperon in the
$p\pi^0; \pi^0 \to \gamma\gamma$ mode and normalize the data sample with 
$\Xi^0 \to \Lambda \pi^0; \Lambda \to p \pi^-$.  Backgrounds are from $\Xi^0$, 
$\Lambda$ and \pccc~ decays; for all decays other than 
$\Xi^0 \to \Lambda \pi^0; \Lambda \to p \pi^-$ and \pccc, Monte Carlo samples 
corresponding to 10 or more times the data sample have been generated.  The kaon
background is studied with data events where the highest-momentum track is 
negatively charged.  For \pccc, high momentum $\pi^+$ and $\pi^-$ are equally
probable, but the hyperon signal is overwhelmingly comprised of events with 
high-momentum positively charged tracks.  Background from 
$\Xi^0 \to \Lambda \pi^0; \Lambda \to p \pi^-$ is suppressed by relying on the
neutrino in the signal mode to produce a missing momentum component perpendicular
to the line of $\Xi^0$ flight.  The overall background level is very low, and
there are nine events in the data, as shown in Figure 1.  We obtain, as a 
preliminary result, Br(\hyp) $= (4.3 \pm1.4) \times 10^{-6}$.

\begin{figure}
\begin{center}
\psfig{figure=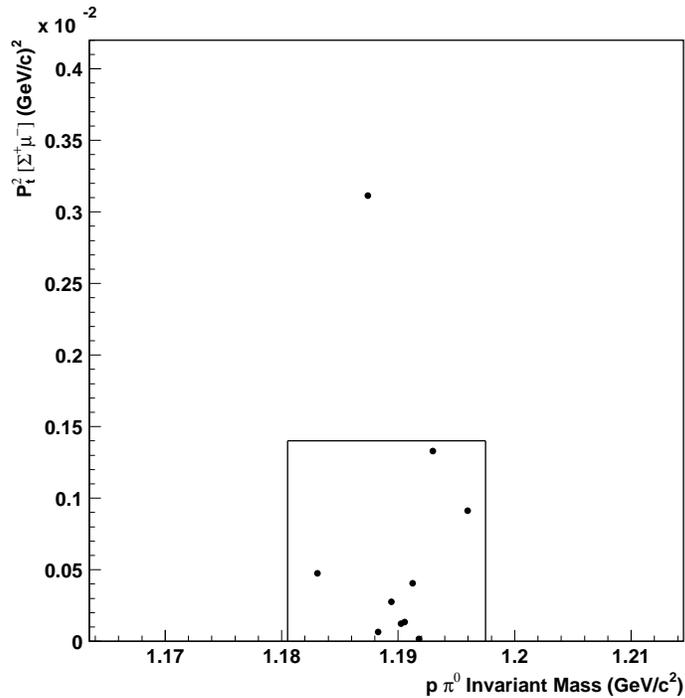,width=10cm}
\caption{Reconstructed $P_\perp$ {\it vs.} $M_{p\pi^0}$ for \hyp candidates
         in the 1999 KTeV data.
\label{fig:errs}}
\end{center}
\end{figure}

\section{\Vus}
\label{sec:three}

A long standing issue in flavor physics has been a discrepancy, at the 
2$\sigma$ level, of the measured values of $|V_{ud}|$, $|V_{us}|$ and 
$|V_{ub}|$ from the unitarity constraint for the first row of the CKM matrix. 
It should be noted that the first row of the matrix is the one that provides
the most stringent test of 3-generation unitarity.
The value of $|V_{ud}|$ is precisely known from nuclear and neutron beta decays,
and $|V_{ub}|$ is small enough to be irrelevant here.  \Vus~ may be
determined from semileptonic kaon decays; in the past only the \Ke~
mode has been used, due to uncertainty in the form factors of the \Km~ mode.

To address this discrepancy, KTeV has sought improved measurements of 
$\Gamma$(\Ke) and $\Gamma$(\Km).  There being no other decay modes that are 
known to sufficient precision to normalize the data set, we have measured 
five ratios of partial widths for the modes \Ke, \Km, \pccc, \pnnn, 
\cc and \nn.  These modes account for nearly all $K_L^0$ decays, and this 
fact may be used to extract the branching ratios for the semileptonic modes.
With the previously accepted~\cite{wrong} value (see the presentation of Gaia 
Lanfranchi at this conference) for the lifetime of the $K_L^0$, partial widths 
for the semileptonic modes and then values for \Vus~ have been extracted.

A full exposition of this work and the related analyses spans a number of publications:
\begin{itemize}
   \item  The radiative corrections - reference~\cite{Andre}
   \item  Measurement of form factors for \Ke~ and \Km~ - reference~\cite{form}
   \item  Check with radiative \Ke~ and \Km~ decays - reference~\cite{rad}
   \item  The measurement of the partial width ratios - reference~\cite{rats}
   \item  The extraction of \Vus~ from these - reference~\cite{voila}
\end{itemize}

The final result, using $f_+(0) = 0.961\pm0.008$, is
\Vus~ $ = 0.2252 \pm0.0008_{\mathrm {KTeV}} \pm0.0021_{\mathrm {ext}}$,
where the ${\mathrm {KTeV}}$ uncertainty includes uncertainties in the KTeV
branching fractions and form factor measurements, and the ${\mathrm {ext}}$ 
uncertainty includes uncertainties in $f_+(0)$, $K_L^0$ lifetime, and radiative 
corrections.  This result does resolve the unitarity discrepancy, and the 
attendant results show a high degree of internal consistency.  While our 
branching ratios are not in good agreement with the values listed in the 2002 
PDG, many of the discrepancies may be explained by postulating that 
Br(\Ke) in the 2002 PDG was too low.  In discussion I emphasized that, as noted 
by Cirigliano {\it et.al.}~\cite{right}, the PDG did not actually have a precise 
and direct measurement of Br(\Ke) on hand.  The value that they recommended was 
inferred from reports of other measurements and the indisputable constraint that 
the sum of the branching ratios must be one.  It was, if you would, a "global 
fit", albeit a substantially simpler analysis than many calculations going by 
that name.  Caveat!

\section{Acknowledgements}
\label{sec:four}

I very much want to thank the organizers of this most excellent conference, 
and in particular the gracious and highly competent Elizabeth Hautefeuille 
as well as the sagacious and justly renowned Jean Tran Thanh Van.

The KTeV collaboration gratefully acknowledges the support and effort of the
Fermilab staff and the technical staffs of the participating institutions for
their vital contributions.  This work was supported in part by the U.S. Dept. of
Energy, the National Science Foundation, and the Ministry of Education and Science
of Japan.

\end{document}